\documentclass[12pt]{article}
\usepackage{graphicx}
\usepackage{bm}

\newcommand\pubnumber{CMS CR-2014/379}
\newcommand\pubdate{November 14, 2014}

\def\zurich{Physik-Institut der Universit\"at Z\"urich\\
Winterthurerstrasse 190, 8057 Zurich, Switzerland}

\def\Title#1{\begin{center} {\Large #1 } \end{center}}
\def\Author#1{\begin{center}{ \sc #1} \end{center}}
\def\Address#1{\begin{center}{ \it #1} \end{center}}

\newcommand\pubblock{\rightline{\begin{tabular}{l} \pubnumber\\
         \pubdate  \end{tabular}}}
\newenvironment{Abstract}{\begin{quotation}  }{\end{quotation}}
\newenvironment{Presented}{\begin{quotation} \begin{center} 
             PRESENTED AT\end{center}\bigskip 
      \begin{center}\begin{large}}{\end{large}\end{center} \end{quotation}}

\begin{document}
\begin{titlepage}
\pubblock

\vfill
\Title{Beyond Standard Model Higgs}
\vfill
\Author{Clemens Lange, on behalf of the ATLAS and CMS collaborations}
\Address{\zurich}
\vfill
\begin{Abstract}
Recent LHC highlights of searches for Higgs bosons beyond the Standard Model are presented. The results by the ATLAS and CMS collaborations are based on 2011 and 2012 proton-proton collision data at centre-of-mass energies of 7 and 8~TeV, respectively. They test a wide range of theoretical models.
\end{Abstract}
\vfill
\begin{Presented}
XXXIV Physics in Collision Symposium \\
Bloomington, Indiana,  September 16--20, 2014
\end{Presented}
\vfill
\end{titlepage}
\def\thefootnote{\fnsymbol{footnote}}
\setcounter{footnote}{0}

\section{Introduction}

The discovery of a Higgs boson with a mass of around 125~GeV by the ATLAS and CMS collaborations \cite{Aad:2012tfa,Chatrchyan:2012ufa} was a great success for elementary particle physics. However, it remains to be seen whether the observed state, which is compatible with the Standard Model (SM) Higgs, really is the SM Higgs, or if there are additional Higgs bosons. Moreover, the discovery of the SM Higgs still would not explain quadratically divergent self-energy corrections at high energies, the so-called hierarchy problem, and would also leave further questions such as the origin of dark matter unanswered. The SM is therefore most likely incomplete. Analyses of the Higgs couplings based on the available experimental data can only set a limit on the Higgs branching ratio to Beyond Standard Model (BSM) particles of $\mathrm{BR}_\mathrm{BSM} < 32\%$ at 95\% C.L.\ assuming no modification at tree level \cite{CMS-PAS-HIG-14-009,ATLAS-CONF-2014-010}. There is plenty of room for Higgs physics beyond the Standard Model.

There are several theoretical models with an extended Higgs sector. Among the most popular and generic ones are the so-called 2 Higgs Doublet Models (2HDM), see e.g.\ Ref.~\cite{Branco:2011iw} for a detailed review. Having two complex scalar SU(2) doublets, they are an effective extension of the SM. In total, five different Higgs bosons are predicted: two CP even and one CP odd electrically neutral Higgs bosons, denoted by $h$, $H$, and $A$, respectively, and two charged Higgs bosons, $H^\pm$. The parameters used to describe a 2HDM are the Higgs boson masses, $m_i$, and the ratios of their vacuum expectation values, $\tan \beta$. There are several classes of 2HDMs, the most relevant ones for this review are the so-called type I, where one SU(2) doublet gives masses to all leptons and quarks and the other doublet essentially decouples from fermions, and the so-called type II, where one doublet gives mass to up-like quarks (and potentially neutrinos) and the down-type quarks and leptons receive mass from the other doublet.
Another intriguing theoretical model is supersymmetry (SUSY), which brings along super-partners of the SM particles. The Minimal Supersymmetric Standard Model (MSSM), whose Higgs sector is equivalent to the one of a constrained 2HDM of type II and the next-to MSSM (NMSSM) are among the experimentally best tested models, because they provide good benchmarks for SUSY. ATLAS and CMS also perform general searches for new heavy particles that can decay to a pair of Higgs bosons, which are motivated by (warped) extra dimension models. The appealing property of all the models described above is that they each solve at least one of the problems the SM does not explain.

\section{Results}

Due to the large number of results, this review can only point out highlights and does not claim to be complete. Instead, the most recent and sensitive analyses are presented. Limits shown in the following are set at a confidence level (C.L.) of 95\% if not stated otherwise.

\subsection{Heavy neutral Higgs bosons decaying to a pair of $\bm{\tau}$ leptons}
\label{sec:MSSMHtautau}

When searching for a Higgs boson of unknown mass, the decay to a pair of $\tau$ leptons provides a good compromise between Higgs branching ratio and expected background processes. In their search for heavy neutral Higgs bosons, generally denoted by $\phi$, the ATLAS and CMS experiments exploit most of the $\tau$ leptons' decay channels \cite{Khachatryan:2014wca,ATLAS-CONF-2014-049}. To account for the different possible production channels, two analysis categories are used: one for events that have at least a b-tagged jet (associated production) and another where no b-tagged jets are present (gluon fusion). One of the major challenges of this analysis is the reconstruction of the mass of the the $\tau$ lepton pair. Both experiments apply slightly different maximum likelihood based methods using the visible $\tau$ decay products and the transverse missing momentum. For the fully hadronic final state ATLAS exploits the di-$\tau$ transverse mass.

\begin{figure}[htb]
\centering
\includegraphics[width=0.5\textwidth]{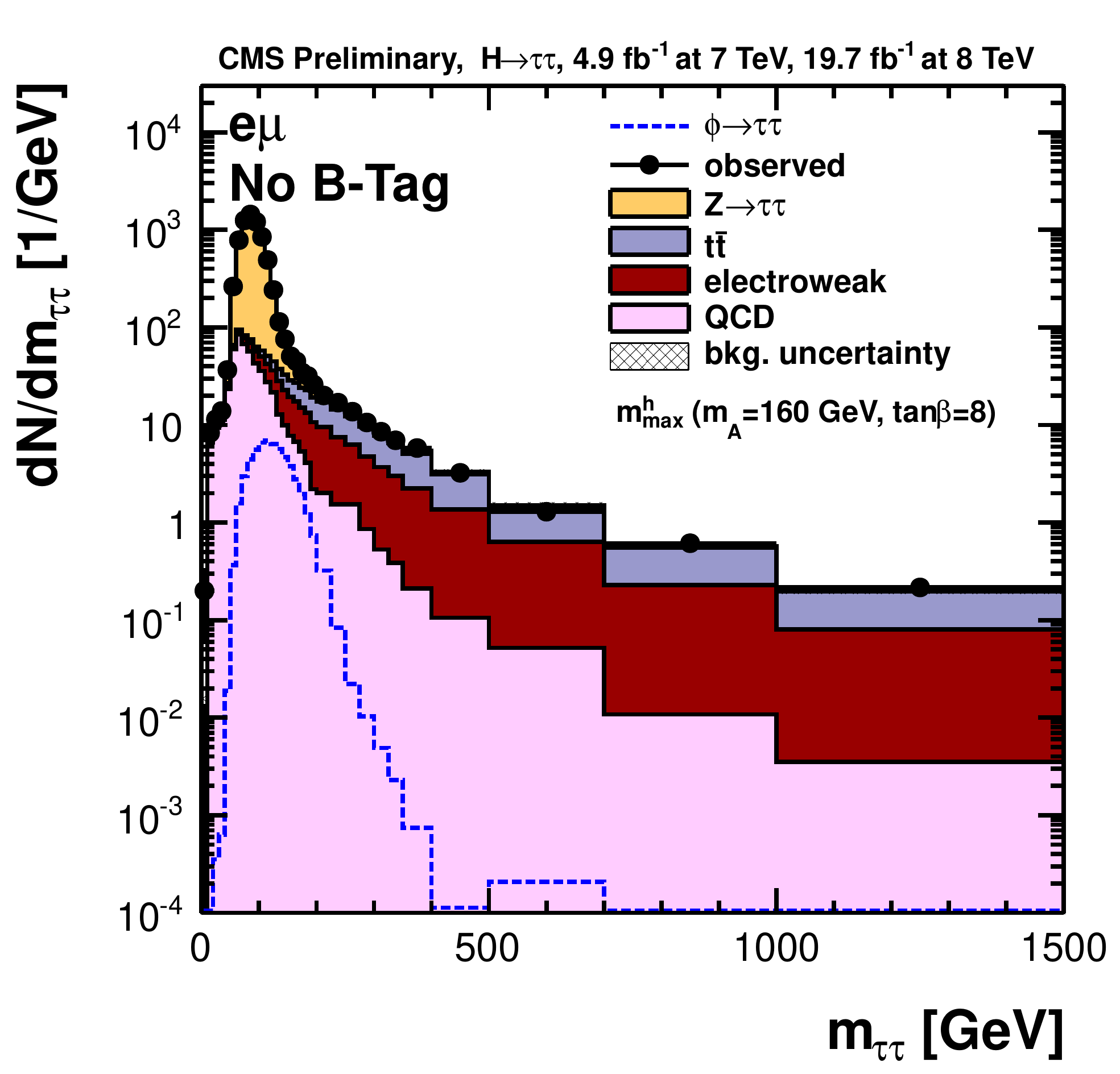}
\includegraphics[width=0.48\textwidth]{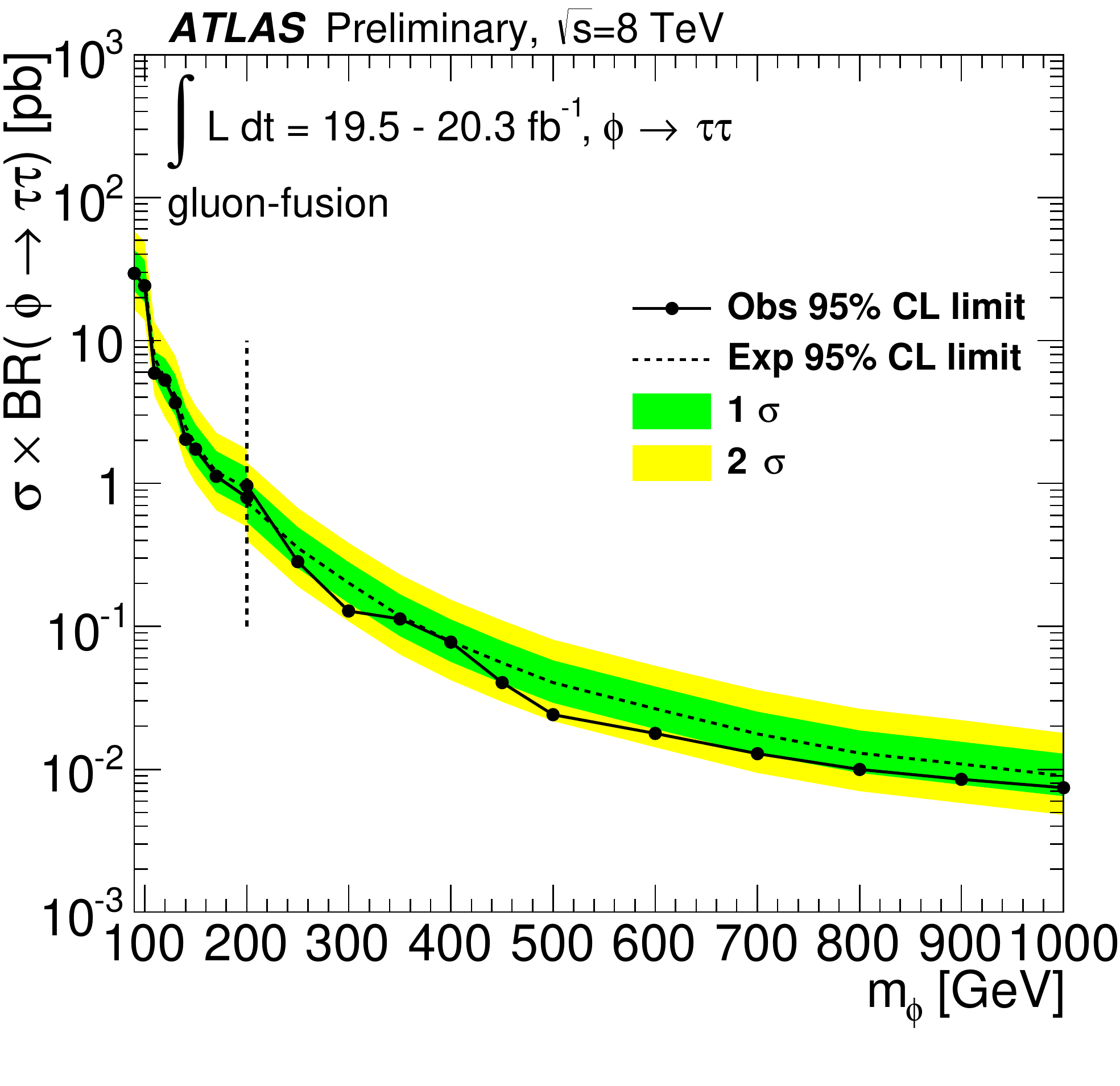}
\caption{Reconstructed di-$\tau$ mass in the no-b-tag category for the $e \mu$ channel for CMS \cite{Khachatryan:2014wca} (left), and expected and observed $\sigma \times \mathrm{BR}( \phi \rightarrow \tau\tau)$ limits for the gluon fusion production channel as a function of the Higgs mass, $m_\phi$, for ATLAS \cite{ATLAS-CONF-2014-049} (right).}
\label{fig:mtautau}
\end{figure}

The dominant background for this analysis are $Z \rightarrow \tau\tau$ events. They are estimated using a so-called embedding technique, where in $Z \rightarrow \mu\mu$ data events the muons are replaced by simulated $\tau$ decay products. As shown in Fig.~\ref{fig:mtautau} good agreement of data with simulation is achieved and all distributions are well-described by the background-only hypothesis. Therefore, cross section times branching ratio limits are calculated and presented separately for the two production mechanisms, which are also shown in Fig.~\ref{fig:mtautau}. The results are additionally interpreted in several MSSM scenarios, which take the Higgs boson at a mass of 125~GeV into account. It should in particular be noted that the $m_h^{\mathrm{mod+}}$ scenario is better suited for this case than the previously used $m_h^{\mathrm{max}}$ scenario. Very low upper limits on $\tan\beta$ are obtained for small Higgs boson masses.

\subsection{Charged Higgs bosons}

\begin{figure}[htb]
\centering
\includegraphics[width=0.48\textwidth]{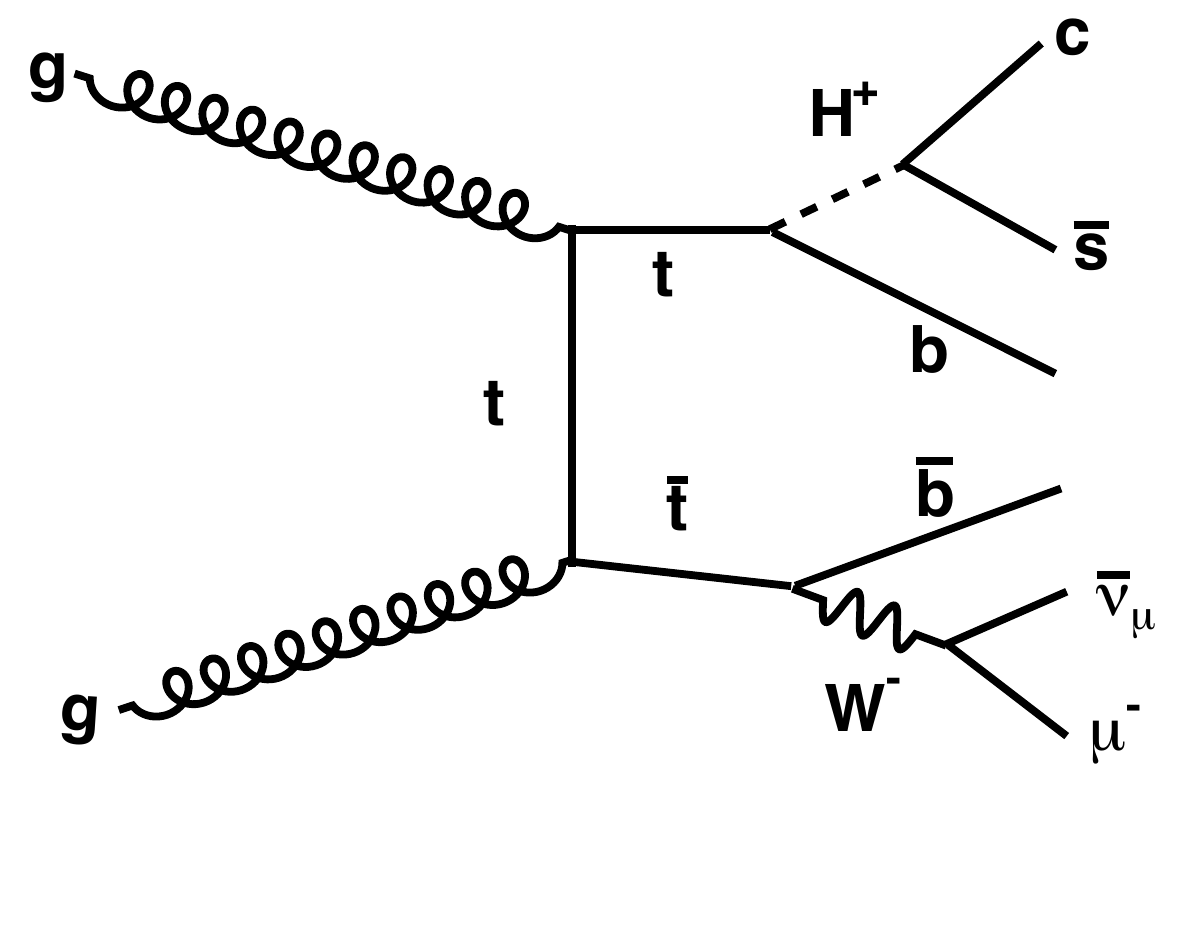}
\includegraphics[width=0.48\textwidth]{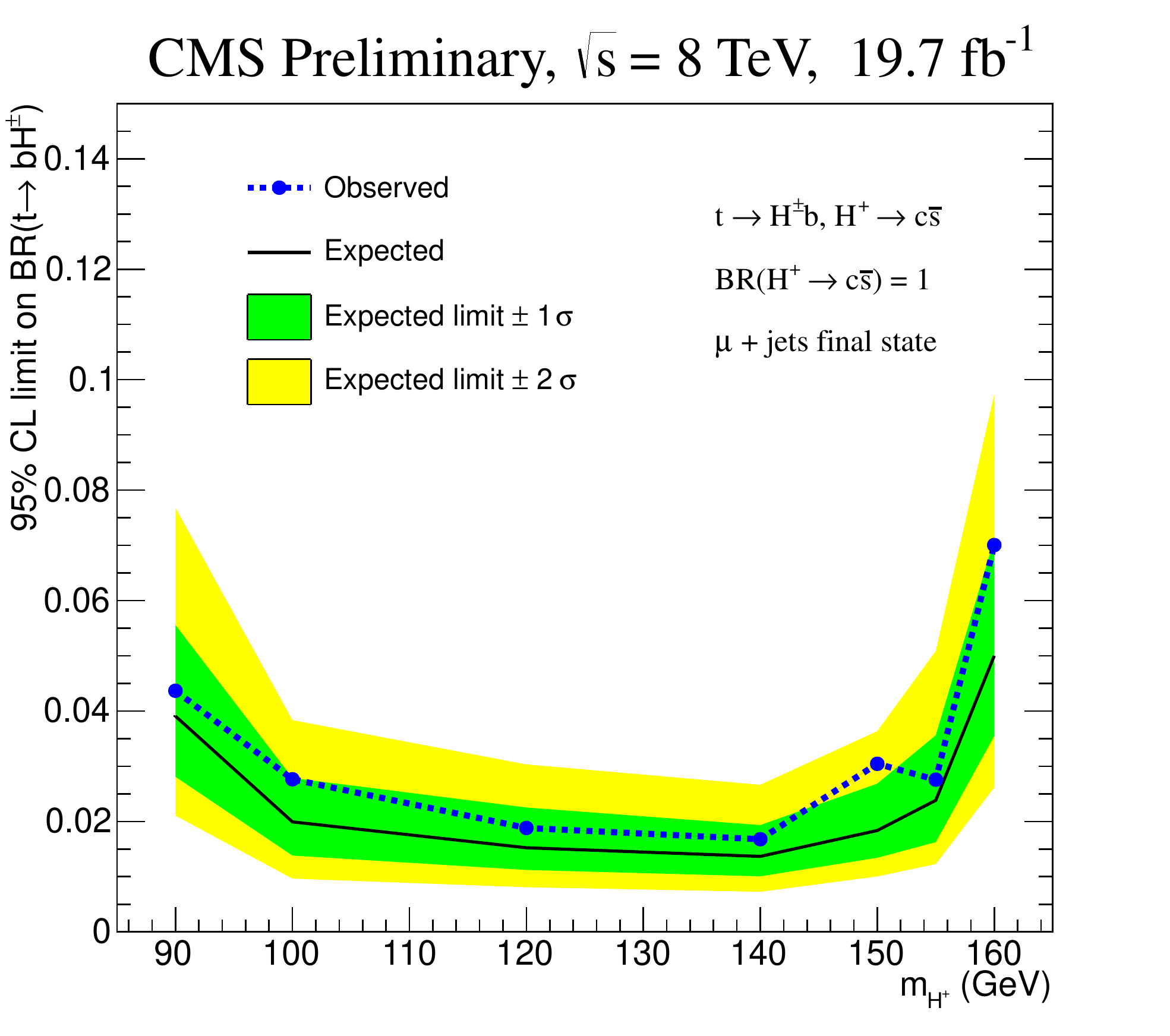}
\caption{Feynman diagram of the production and decay topology of a charged Higgs boson (left), and expected and observed BR($t \rightarrow H^+ b$) limits assuming BR$(H^+ \rightarrow c\bar{s})=1$ as a function of the Higgs mass, $m_{H^+}$, for CMS \cite{CMS-PAS-HIG-13-035} (right).}
\label{fig:chargedH}
\end{figure}

In the search for charged Higgs bosons, the relation between the mass of the Higgs boson and the top quark mass dictates both the production modes and the decay channels. In the analysis presented here \cite{CMS-PAS-HIG-13-035,Aad:2013hla}, the charged Higgs boson is assumed to be lighter than the top quark. This means that for reasonably small values of $\tan \beta$ the dominant decay for a positively charged Higgs boson is to a $c\bar{s}$ pair (other searches have been performed on 7 TeV data in the $H^+ \rightarrow \tau \nu$ channel \cite{Aad:2012tj}, which is the dominant decay for larger values of $\tan \beta$). As a consequence, these charged Higgs events will have the same topology as top quark pair events as shown in Fig.~\ref{fig:chargedH}. One therefore searches for a second peak in the di-jet mass spectrum for which all non b-tagged jets are taken into account. A kinematic fit is used to constrain both top quark candidates to $m_t = 172.5$~GeV. Limits on the branching ratio BR($t \rightarrow H^+ b$) are determined assuming a 100\% branching ratio of $H^+ \rightarrow c\bar{s}$, which are shown in Fig.~\ref{fig:chargedH}. The analysis is in particular sensitive in the charged Higgs mass range of 90--155~GeV. The limits set are applicable to any BSM resonance with corresponding production and decay topology.

\subsection{Heavy neutral Higgs bosons decaying to a pair of lighter neutral Higgs bosons}

The SM rate of Higgs pair production is very small, but resonant pair production is motivated by several BSM models such as decays of heavy (N)MSSM Higgs bosons as well as radion and Kaluza-Klein excitations of gravitons in the context of warped extra dimensions. This usually goes along with an enhanced production cross section meaning that analyses are already sensitive to these models with the current dataset. The $b\bar{b} \gamma \gamma$ final state is particularly intriguing because one can exploit the high branching ratio of $H \rightarrow b\bar{b}$ and the high mass resolution of $H \rightarrow \gamma \gamma$.

\begin{figure}[htb]
\centering
\includegraphics[width=0.48\textwidth]{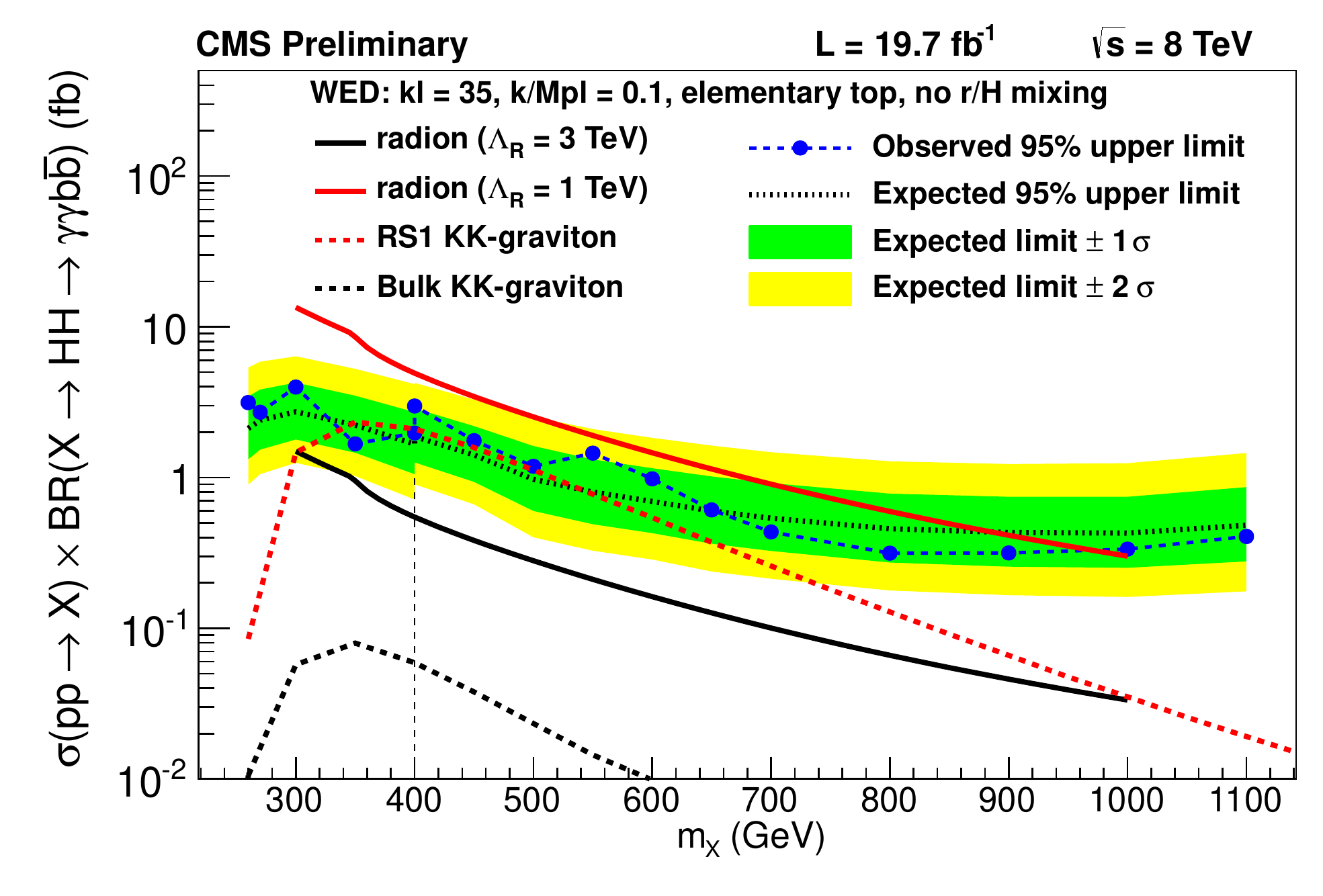}
\includegraphics[width=0.48\textwidth]{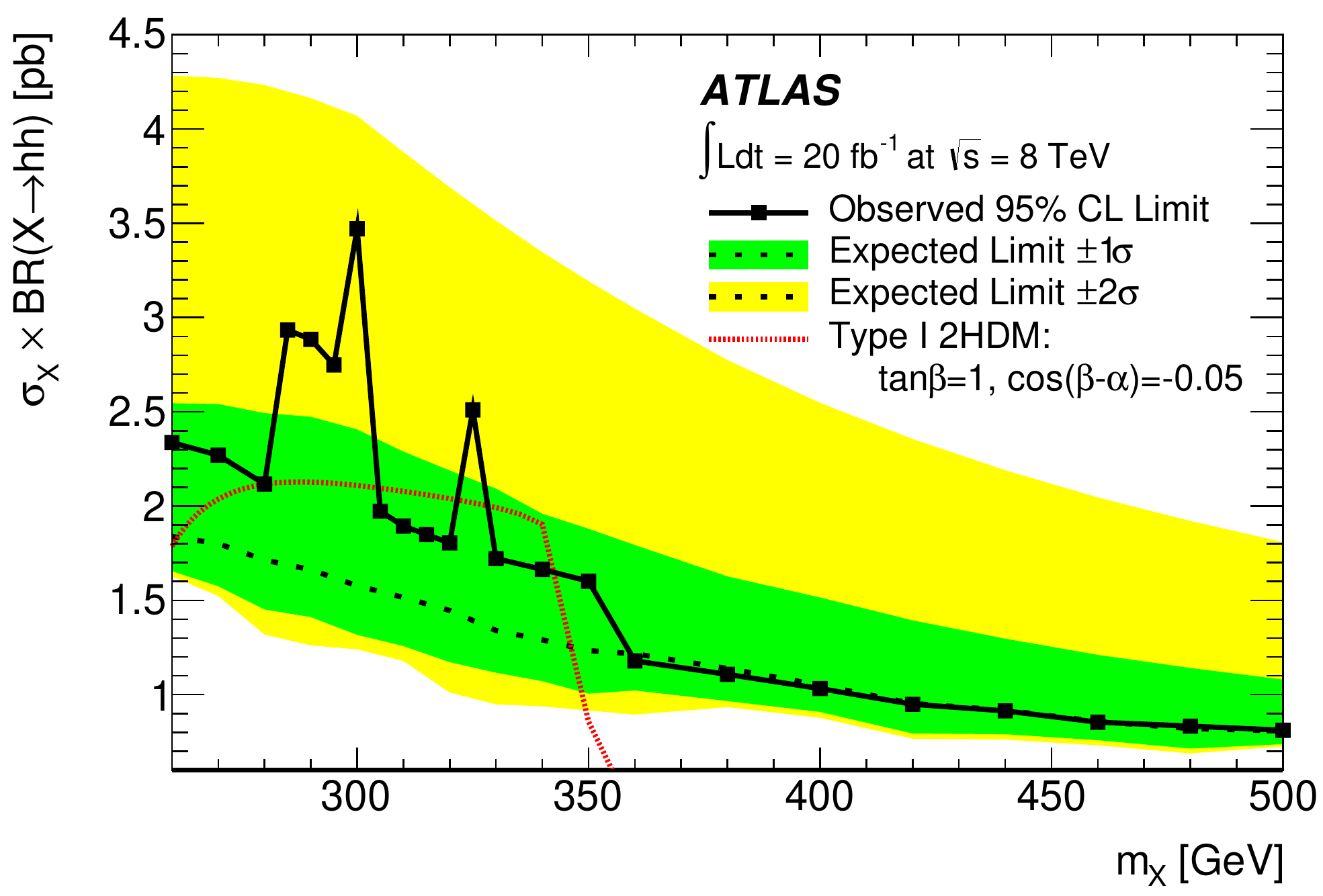}
\caption{Expected and observed $\sigma (pp \rightarrow X) \times \mathrm{BR}(X \rightarrow b\bar{b} \gamma \gamma)$ limits as a function of the resonance mass, $m_{X}$, for CMS in comparison with theoretical cross section predictions \cite{CMS-PAS-HIG-13-032} (left), and expected and observed $\sigma (pp \rightarrow X) \times \mathrm{BR}(X \rightarrow hh)$ limits as a function of the resonance mass, $m_{X}$, for ATLAS \cite{Aad:2014yja} (right).}
\label{fig:bbgg}
\end{figure}

The event selection is based on the SM Higgs analyses in the respective channels. In order to suppress the SM continuum of $b\bar{b}$, a mass constraint in the case of CMS \cite{CMS-PAS-HIG-13-032} and a mass window cut in the case of ATLAS \cite{Aad:2014yja} are applied making use of the known Higgs mass of 125~GeV. The resonant searches do not show any deviation from the SM expectation. Therefore, upper limits are set which exclude radions with masses less than 970~GeV and certain Kaluza-Klein gravitons in the mass range of 340--400~GeV as shown in Fig.~\ref{fig:bbgg}. The ATLAS analysis additionally searches for non-resonant $hh$ production where a slight excess of $2.4\sigma$ is observed, which is for example compatible with a 2HDM of type I.

In addition to the $b\bar{b} \gamma \gamma$ final state, this narrow resonance search is also performed in the $b\bar{b} b\bar{b}$ channel by both ATLAS and CMS \cite{CMS-PAS-HIG-14-013,ATLAS-CONF-2014-005}. Due to the lower mass resolution a window of $\pm 35$~GeV around the Higgs mass is used. The dominant QCD multi-jet background is fit using an analytic function, which is validated in sideband regions. No statistically significant signal is observed. Hence limits are set in a mass range up to 1400~GeV.

\subsection{Lepton flavour violating Higgs decays}

\begin{figure}[htb]
\centering
\includegraphics[width=0.48\textwidth]{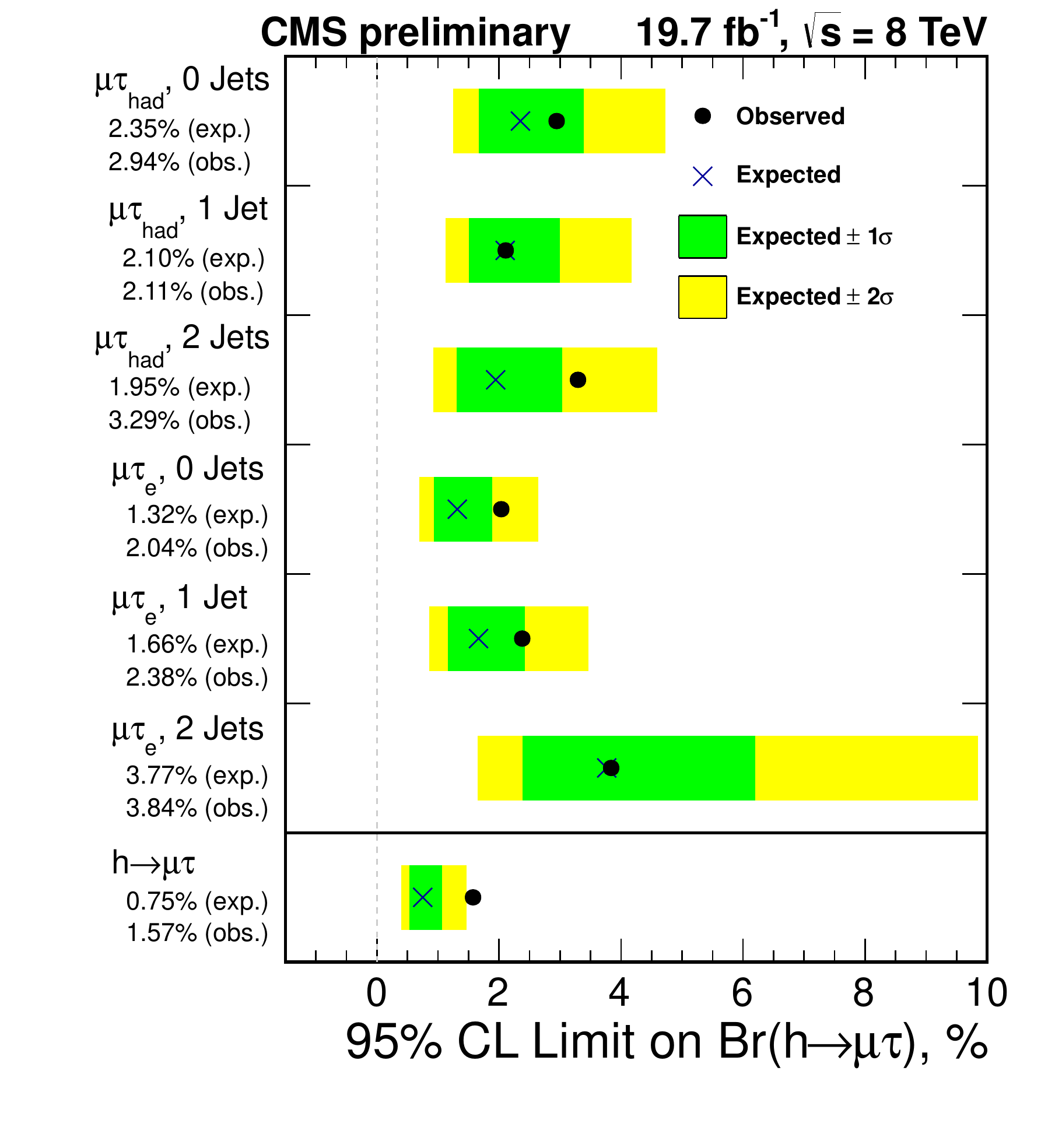}
\includegraphics[width=0.48\textwidth]{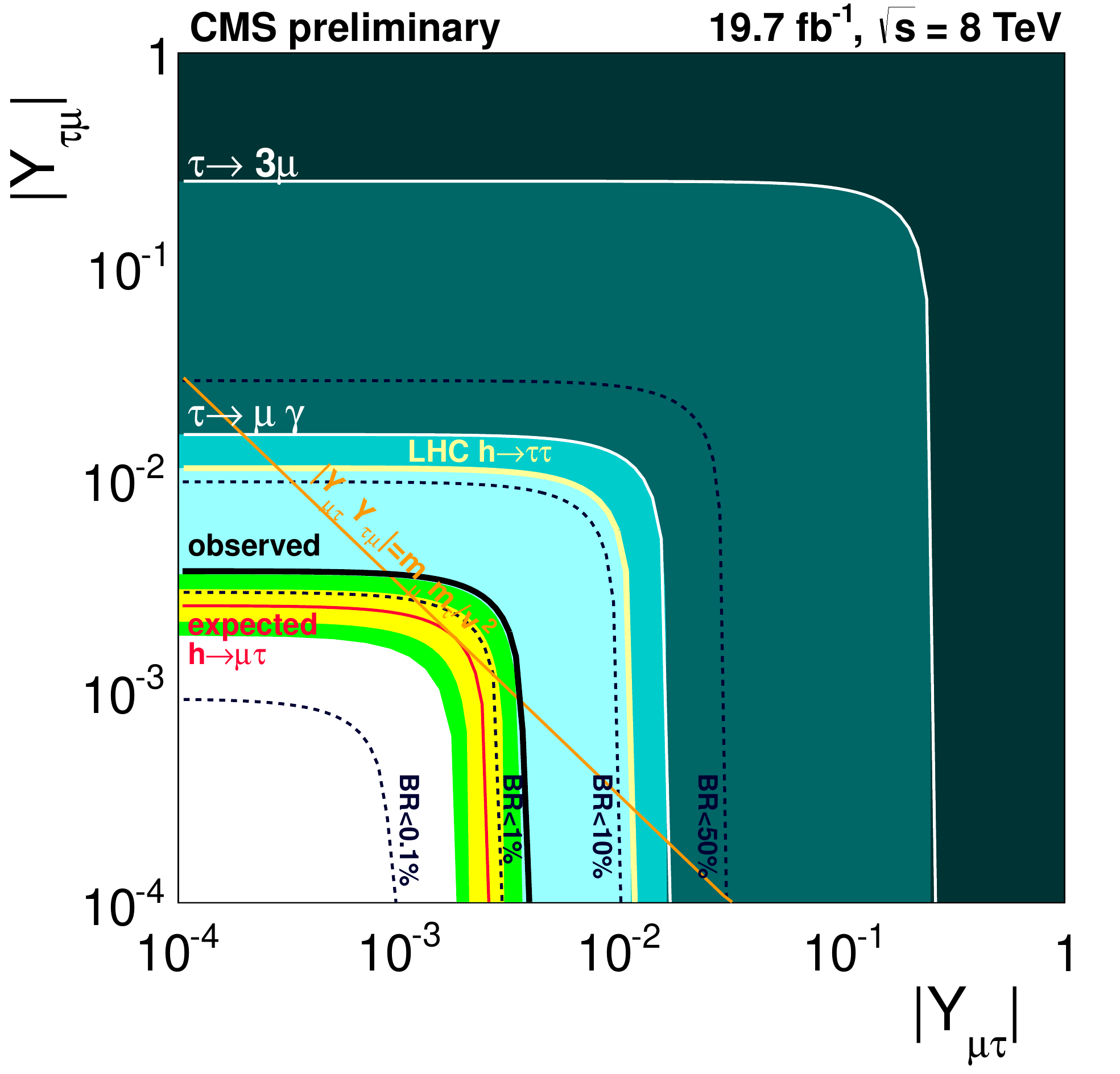}
\caption{Expected and observed BR$(H \rightarrow \mu \tau)$ limits for the different analysis channels and their combination (left), and constraints on the flavour violating Yukawa couplings, $|Y_{\mu\tau}|$, $|Y_{\tau\mu}|$, in comparison to other analyses \cite{CMS-PAS-HIG-14-005} (right).}
\label{fig:flavvio}
\end{figure}

In 2HDM and Randall-Sundrum models lepton flavour violating Higgs decays can occur. The analysis presented in Ref.~\cite{CMS-PAS-HIG-14-005} searches for a mass resonance at approximately 125~GeV in the process $H \rightarrow \mu \tau$ decays in the $\mu \tau_e$ and the $\mu \tau_\mathrm{had}$ channels. This analysis has a similar signature as the $\phi \rightarrow \tau\tau$ analysis in Sec.~\ref{sec:MSSMHtautau}, but different kinematics. The invariant mass of the $\tau$ decay products is approximated using the collinear mass. Combining the different analysis channels an upper limit of BR$(H \rightarrow \mu \tau)$ of about 1.5\% is set as shown in Fig.~\ref{fig:flavvio}. The best fit yields BR$(H \rightarrow \mu \tau) = (0.89 \pm 0.39)\%$, i.e. a small excess of $2.5\sigma$, which is, however, still compatible with zero. The sensitivity of the search
is an order of magnitude better than the existing indirect limits and also sets the best limits on flavour-violating $\tau\mu$ Yukawa couplings to date as shown in Fig.~\ref{fig:flavvio}.

\subsection{Invisible Higgs decays}

\begin{figure}[htb]
\centering
\includegraphics[width=0.72\textwidth]{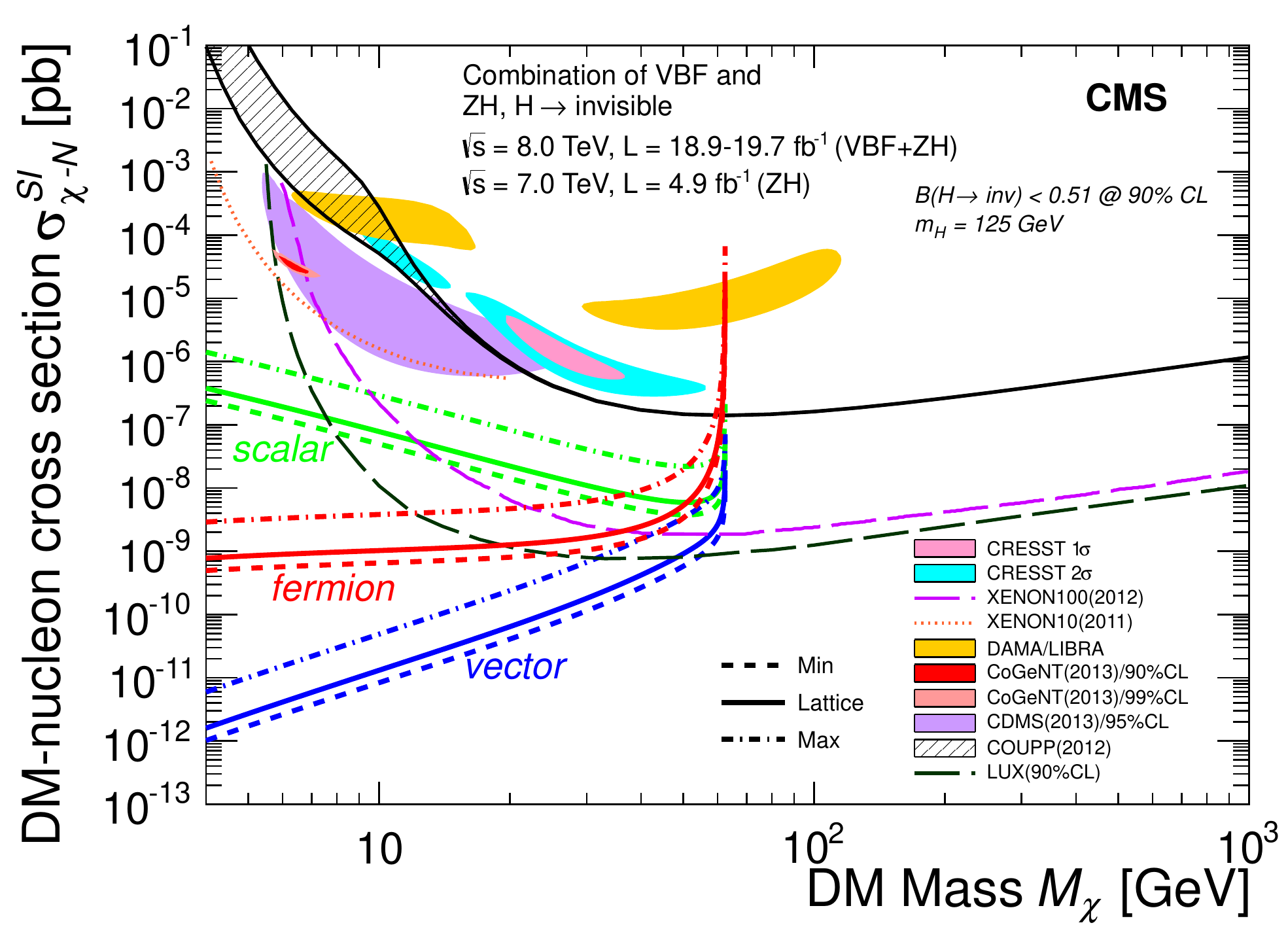}
\caption{Upper limits on the spin-independent DM-nucleon cross section $\sigma^\mathrm{SI}_{\chi-N}$ in Higgs-portal models, derived for $m_H = 125$~GeV and $\mathrm{BR}(H \rightarrow \mathrm{invisible}) < 0.51$ at 90\% C.L., as a function of the DM mass. Limits are shown separately for scalar, vector and fermion DM. The solid lines represent the central value of the Higgs-nucleon coupling, which enters as a parameter, and is taken from a lattice calculation, while the dashed and dot-dashed lines represent lower and upper bounds on this parameter. Other experimental results are shown for comparison \cite{Chatrchyan:2014tja}.}
\label{fig:darkmatter}
\end{figure}

The decay of a Higgs boson to invisible particles is predicted in several New Physics models such as the decay to neutralinos in SUSY models and the decay to graviscalars in extra dimension models. In order to be able to identify such events, the Higgs boson needs to be produced with visible particles as is the case in associated $ZH$ production or vector boson fusion (VBF), where the latter features a higher cross section. The event signature consists therefore either of two forwards jets (VBF) or a pair of leptons ($ZH$) and large missing transverse energy. The main background processes are vector bosons ($V$) where the boson decays unseen, either produced as $V$+jets in the case of the VBF signal and $VV$ diboson production in case of the $ZH$ signal search. The $V$+jets contribution is estimated using visible $V$ decay modes where the decay products are removed \cite{Chatrchyan:2014tja,Aad:2014iia}. The data show good agreement with SM backgrounds only.
Limits are set on the $\sigma_{ZH} \times \mathrm{BR}(H \rightarrow \mathrm{invisible})$ based on the missing transverse energy (MET) distribution in case of the ATLAS analysis. The CMS analysis combines the $ZH$ and VBF channels based on the MET and di-jet mass distributions, respectively. The observed limit on the branching ratio is $\mathrm{BR}(H \rightarrow \mathrm{invisible}) < 58\%$ (44\% expected) for a SM Higgs with a mass of 125 GeV. These analyses in particular show significant improvement in sensitivity with respect to earlier direct searches. Therefore, both experiments additionally perform a dark matter (DM) interpretation using the Higgs-portal model. This model proposes a hidden sector with stable dark matter particles, which, if their mass is below $m_H/2$, would contribute to the invisible decay width of the Higgs boson. The analyses are complementary to direct detection of dark matter and sensitive to the dark matter nucleon cross section. The CMS limits set on the DM nucleon cross section as a function of the dark matter candidate mass in comparison with direct searches are shown in Fig.~\ref{fig:darkmatter}.

\subsection{Higgs bosons in long-lived particle searches}

Models such as "split SUSY" or "hidden valleys" propose the existence of long-lived particles. One of these particles could be a neutral Higgs boson, which might be created in the LHC collisions, see References \cite{CMS-PAS-EXO-12-037,Aad:2014yea,ATLAS-CONF-2014-041} for more details. If this was the case, one would observe delayed decays in the detector. The searches therefore focus on leptons with large transverse decay lengths, jets that are only created in the hadronic calorimeter and so-called lepton-jets, i.e.\ collimated jets of leptons and hadrons. These kind of analyses are particularly challenging because the reconstruction algorithms of the experiments need to be adjusted. Furthermore, dedicated triggers need to be put in place to make sure to catch such events. However, as of today no signal has been found. Therefore, limits on the lifetime of the proposed new particles convoluted with branching ratios and cross sections are set.

\section{Summary}

The ATLAS and CMS experiments at the LHC conduct numerous studies reaching beyond the Standard Model Higgs to determine whether the newly discovered state is the SM boson or part of an extended Higgs sector. Any finding in conflict with the SM Higgs would be a hint for New Physics. Several new results based on LHC Run I data have recently been presented, where each analysis faces different challenges. With the upcoming LHC Run II at a higher centre-of-mass energy the reach of these analyses will be extended.

\end{document}